\documentclass[prd,twocolumn,floatfix,nofootinbib,amsmath,amssymb,floatfix]{revtex4}
\usepackage{graphicx,color,dcolumn,booktabs,bm,mathrsfs}
\usepackage{longtable,lscape}
\usepackage{txfonts}
\usepackage{overpic}
\usepackage{amssymb}
\usepackage{indentfirst}
\usepackage{feynmf}   
\usepackage{slashed}  
\usepackage{cases}
\usepackage{color}
\usepackage{multirow}
\usepackage{epstopdf}
\usepackage{CJK}
\usepackage[colorlinks,
            citecolor=blue,
            anchorcolor=red,
            menucolor=red,
            linkcolor=red,
            filecolor=red,
            runcolor=red,
            urlcolor=blue,
            frenchlinks=red]{hyperref}
\usepackage{subfigure}

\begin{document}

\title{The assignments of the bottom mesons within the screened potential model and $^3P_0$ model}

\author{Xue-Chao Feng}
\affiliation{College of Physics and Electronic Engineering, Zhengzhou 
University of Light Industry, Zhengzhou 450002, China}

\author{Wei Hao}
\affiliation{CAS Key Laboratory of Theoretical Physics, Institute of Theoretical Physics, Chinese Academy of Sciences, Beijing 100190,China}
\affiliation{University of Chinese Academy of Sciences (UCAS), Beijing 100049, China}

\author{Li-Juan Liu}\email{liulijuan@zzu.edu.cn}
\affiliation{School of Physics and Microelectronics, Zhengzhou University, Zhengzhou, Henan 450001, China}

\begin{abstract}
In this work, we calculate the mass spectrum of the bottom mesons with a modified nonrelativistic quark model by involving the screening effect, and explore their strong decay properties within the $^3P_0$ model. 
Our results suggest that the $B_1(5721)$, $B^*_2(5747)$, $B_J(5840)$, and $B_J(5970)$ could be reasonably assigned as the $B_1^\prime(1P)$, $B(1^3P_2)$, $B(2^1S_0)$, and $B(1^3D_3)$ respectively. 
The more precise measurements of the excited bottom mesons are crucial to confirm these assignments. 
\end{abstract}

\maketitle

\section{Introduction}

Meson spectroscopy is one of the important subjects in hadron physics, and the heavy-light system offers an excellent laboratory for testing the heavy quark symmetry.  
As we known, most of the charmed and charmed-strange mesons are well established, although there exists some exotic explanations for some states, for example $D^*_{s0}(2317)$ and $D_{s1}(2460)$~\cite{Chen:2016spr,Du:2017zvv}. 
For the family of bottom mesons, the first bottom meson $B$ was observed in 1983 by the CLEO Collaboration~\cite{CLEO:1983mma}, and there are only six excited bottom mesons observed experimentally so far, which are $B^*$, $B_1(5721)$, $B^*_J(5732)$, $B_J(5840)$, $B_J(5970)$, and $B^*_2(5747)$~\cite{PDG2021}. Among those bottom states, only $B$ and $B^*$ are well assigned as the $S$-wave doublet ($1^1S_0$ and $1^3S_1$).

The $B^*_J(5732)$, as the first orbitally excited bottom meson, was observed in 1994 by the OPAL detector at LEP~\cite{OPAL:1994hqv}, and this state has not yet confirmed in the last two decadeds~\cite{PDG2021}. It should be streesed that its signal can be interpreted as stemming from several narrow and broad resonances, as discussed in Review of Particle Physics (RPP)~\cite{PDG2021}, thus we will not discuss this state. In 2007, the D0 Collaboration reported two narrow orbitally excited ($L=1$) bottom mesons,  $B_1(5721)$ with $J^P=1^+$ and  $B^*_2(5747)$ with $J^P=2^+$~\cite{D0:2007vzd}, later confirmed by the CDF and LHCb Collaborations~\cite{CDF:2008qzb,CDF:2013www,LHCb:2015aaf}. The $B_J(5970)$ was first observed in 2013 by the CDF Collaboration~\cite{CDF:2013www}. In 2015 the LHCb Collaboration reproted two states $B_J(5840)$  and $B_J(5960)$~\cite{LHCb:2015aaf}, where the latter one should be the same state as the $B_J(5970)$ since they have the similar properties. Here we list the masses, widths, and the quantum numbers of $B_1(5721)$, $B^*_J(5732)$, $B_J(5840)$, $B_J(5970)$, and $B^*_2(5747)$ in Table~\ref{experment}.

\begin{table}[!htpb]
\begin{center}
\caption{ \label{experment} Masses, decay widths, and quantum numbers of the bottom mesons~\cite{PDG2021}.}
\footnotesize
\setlength{\tabcolsep}{1mm}{
\begin{tabular}{lcccc}
\hline\hline
  state                 &mass  (MeV)                          &width  (MeV)                & $I(J^{P})$          \\\hline
  $B_J^*(5732)$             &$5698\pm8$                      &$128\pm18$             & $?(?^?)$  \\ 
  $B_1(5721)^0$             &$5726.1\pm 1.3$                 &$27.5\pm3.4$           & $1/2(1^+)$      \\
  $B^*_2(5747)^0$           &$5739.5\pm 0.7$                 &$24.2\pm1.7$           & $1/2(2^+)$ \\
  $B_J(5840)$               &$5863\pm9$                      &$127\pm 40$            & $1/2(?^?)$    \\
  $B_J(5970)$               &$5971\pm5$                      &$81\pm12$              & $1/2(?^?)$     \\
  \hline\hline

\end{tabular}}
\end{center}
\end{table}

Based on the experimental measurements of the bottom mesons, there are many theoretical discussions in literatures~\cite{Zeng:1994vj, Godfrey:1985xj, Lahde:1999ih, DiPierro:2001dwf, Zhong:2008kd, Ebert:2009ua, Godfrey:2016nwn, Lu:2016bbk, Kher:2017mky, Asghar:2018tha, Godfrey:2019cmi, Yu:2019iwm,Chen:2022fye,li:2021hss,Narison:2020wql}. The $B^*_2(5747)$ is commonly regarded as the $B(1^3P_2)$ state~\cite{Ebert:2009ua, Godfrey:2016nwn, Lu:2016bbk, Kher:2017mky, Asghar:2018tha, Godfrey:2019cmi, Yu:2019iwm,Chen:2022fye}, and the $B_1(5721)$ is explained as the $B^\prime(1P_1)$ in Refs.~\cite{Lu:2016bbk,Godfrey:2019cmi,Yu:2019iwm,Chen:2022fye}, $B(1P_1)$ in Refs.~\cite{Kher:2017mky,Asghar:2018tha}.

 Although the LHCb Collaboration has suggested that the $B_J(5840)$ and $B_J(5970)$ could be the $2^1S_0$ and $2^3S_1$ states, respectively~\cite{LHCb:2015aaf}, the mass difference between them is about 110~MeV, much larger than the theoretical predictions of the $2S$  mass splitting, which is 39~MeV of Godfrey-Isgur model~\citep{Godfrey:1985xj}, 30~MeV in alternate relativized (AR) model~\cite{Godfrey:2016nwn}, and 28~MeV in nonrelativistic potential quark model~\cite{Asghar:2018tha}. On the other hand, there are different interpretations for the $B_J(5840)$ and $B_J(5970)$, 
because their masses and widths can not be simultaneously  reasonably reproduced.



 As discussed in Ref.~\cite{bai:2009}, the quenched quark models, incorporating a coulomb term at short distances and the linear confining interaction at large distances, will not be reliable for the high excited mesons. This is because the linear potential, which is expected to be dominant in this mass region, will be screened and softened by the vacuum polarization effects of dynamical fermions. It is shown that the screened potential plays an important role in describing the spectra and decay properties of the charmed, charmed-strange mesons and charmonium~\cite{Song:2015fha,Song:2015nia,Wang:2018rjg}. Thus one expect that the screened potential could improve the description of the bottom mesons.

In this work, we will investigate the possible assignments of the $B_1(5721)$, $B^*_2(5747)$, $B_J(5840)$, and $B_J(5970)$ by employing the modified nonrelativistic quark model with screening effect, and the $^3P_0$ model to calculate the mass spectrum and decay properties of the bottom mesons.

This article is organized as follows. In  Sec.~\ref{sec:model}, we give a brief review about the modified nonrelativistic quark model and the $^3P_0$ model. In Sec.~\ref{sec:result}, we present the numerical results and discuss the possible assignments of the bottom mesons. The summary is given in Sec.~\ref{sec:summary}

\section{Theoretical models}
\label{sec:model}

\subsection{Modified nonrelativistic quark model}
The nonrelativistic quark model~\cite{Lakhina:2006fy}, as one of the successful quark models, is proposed by Lakhina and Swanson, and has already been used to  describe the heavy-light mesons and heavy quarkonium successfully, such as the charmed-strange mesons~\citep{Li:2010vx} and the bottom mesons~\citep{Lu:2016bbk}. In the model, the Hamiltonian of a $q\bar{q}$ meson system is defined as~\citep{Li:2010vx,Lu:2016bbk}
    \begin{equation}
    \label{ha}
    H = H_0+H_{sd}+C_{q\bar{q}}, 
    \end{equation}
where $H_0$ is the zeroth-order Hamiltonian, $H_{sd}$ is the spin-dependent Hamiltonian, and $C_{q\bar{q}}$ is a constant, which will be fixed by experimental data.
The $H_{0}$ can be compressed as
    \begin{eqnarray}
          H_0 &=& \frac{\boldsymbol{p}^2}{M_r}-\frac{4}{3}\frac{{\alpha}_s}{r}+br+\frac{32{\alpha}_s{\sigma}^3 e^{-{\sigma}^2r^2}}{9\sqrt{\pi}m_qm_{\bar{q}}} {\boldsymbol{S}}_{q} \cdot {\boldsymbol{S}}_{\bar{q}}, \label{eq:H0}
     \end{eqnarray}
where the confinement interaction includes the standard Coulomb potential $-4\alpha_s/3r$ and linear scalar potential $br$, and the last term is the hyperfine interaction that could be treated nonperturbatively.
In Eq.~(\ref{eq:H0}), $\boldsymbol{p}$ is quark momentum in the system of $q\bar{q}$ meson, $r=|\vec{r}|$ is the $q\bar{q}$ separation, $M_r=2m_qm_{\bar{q}}/(m_q+m_{\bar{q}})$, $m_q$ ($m_{\bar{q}}$) and $\boldsymbol{S}_{q}$ (${\boldsymbol{S}}_{\bar{q}}$) are the mass and spin of the constituent quark $q$ (antiquark $\bar{q}$), respectively.

The spin-dependent term $H_{sd}$ is,
    \begin{eqnarray}
      H_{sd} &=& \left(\frac{\boldsymbol{S}_{q}}{2m_q^2}+\frac{{\boldsymbol{S}}_{\bar{q}}}{2m_{\bar{q}}^2}\right) \cdot \boldsymbol{L}\,\left(\frac{1}{r}\frac{dV_c}{dr}+\frac{2}{r}\frac{dV_1}{dr}\right)\nonumber\\
      &&+\frac{{\boldsymbol{S}}_+ \cdot \boldsymbol{L}}{m_qm_{\bar{q}}}\left(\frac{1}{r} \frac{dV_2}{r}\right) \nonumber\\
      && +\frac{3{\boldsymbol{S}}_{q} \cdot \hat{\boldsymbol{r}}\,{\boldsymbol{S}}_{\bar{q}} \cdot \hat{\boldsymbol{r}}-{\boldsymbol{S}}_{q} \cdot {\boldsymbol{S}}_{\bar{q}}}{3m_qm_{\bar{q}}}V_3\nonumber\\
      && +\left[\left(\frac{{\boldsymbol{S}}_{q}}{m_q^2}-\frac{{\boldsymbol{S}}_{\bar{q}}}{m_{\bar{q}}^2}\right)+\frac{{\boldsymbol{S}}_-}{m_qm_{\bar{q}}}\right] \cdot \boldsymbol{L} V_4,
\end{eqnarray}
with
\begin{eqnarray}
  V_c &=& -\frac{4}{3}\frac{{\alpha}_s}{r}+br,\nonumber \\
  V_1 &=& -br-\frac{2}{9\pi}\frac{{\alpha}_s^2}{r}\left[9\,{\rm ln}(\sqrt{m_qm_{\bar{q}}}r)+9{\gamma}_E-4\right],\nonumber\\
  V_2 &=& -\frac{4}{3}\frac{{\alpha}_s}{r}-\frac{1}{9\pi}\frac{{\alpha}_s^2}{r}\left[-18\,{\rm ln}(\sqrt{m_qm_{\bar{q}}}r)+54\,{\rm ln}(\mu r) \right.\nonumber\\
  &&\left. +36{\gamma}_E+29\right],\nonumber\\
  V_3 &=& -\frac{4{\alpha}_s}{r^3}-\frac{1}{3\pi}\frac{{\alpha}_s^2}{r^3}\left[-36\,{\rm ln}(\sqrt{m_qm_{\bar{q}}}r)+54\,{\rm ln}(\mu r) \right.\nonumber\\
  &&\left. +18{\gamma}_E+31\right],\nonumber\\
  V_4 &=& \frac{1}{\pi}\frac{{\alpha}_s^2}{r^3}{\rm ln}\left(\frac{m_{\bar{q}}}{m_q}\right),
\end{eqnarray}
where $\boldsymbol{S}_{\pm}={\boldsymbol{S}}_q\pm{\boldsymbol{S}}_{\bar{q}}$, $\boldsymbol{L}$ is the relative orbital angular momentum of the $q\bar{q}$ system. We take Euler constant $\gamma_E=0.5772$, the scalar $\mu=1$~GeV, ${\alpha}_s=0.5$, $b=0.14$~GeV$^2$, $\sigma=1.17$~GeV, $m_u=m_d=0.45$~GeV, and $m_b=4.5$ GeV \citep{Li:2010vx,Lu:2016bbk}.

Because the coupled-channel effects become more important for higher radial and orbital excitations of the heavy-light mesons, some modified models have been proposed by including the screening effect~\cite{bai:2009,Song:2015nia,Song:2015fha}, and widely used to calculate mass spectrum of charmed-strange meson~\cite{Song:2015nia}, charm meson~ \cite{Song:2015fha}, charmonium~\cite{Wang:2019mhs,bai:2009}, and bottomonium~\cite{Wang:2018rjg}. The screening effect was introduced by the following  replacement~\cite{Song:2015nia},
\begin{eqnarray}
br\to V^{\text{scr}}(r)=\frac{b(1-e^{-\beta r})}{\beta}, \label{eq:screened}
\end{eqnarray}
where $V^{\text{scr}}(r)$ behaves like $br$ at short distances and constant $b/\beta$ at large distance\cite{Song:2015nia, Song:2015fha}, $\beta$ is parameter which is used to control the power of  the screening effect.

The spin-orbit term in the $H_{sd}$ can be decomposed into
symmetric part $H_{sym}$ and antisymmetric part $H_{anti}$. These two parts can be written as \cite{Lu:2016bbk}
\begin{eqnarray}
H_{sym} &=& \frac{{\boldsymbol{S}}_+ \cdot {\boldsymbol{L}}}{2}\left[\left(\frac{1}{2m_q^2}+\frac{1}{2m_{\bar{q}}^2}\right) \left(\frac{1}{r}\frac{dV_c}{dr}+\frac{2}{r}\frac{dV_1}{dr}\right)\right. \nonumber \\
&& \left.+\frac{2}{m_qm_{\bar{q}}}\left(\frac{1}{r} \frac{dV_2}{r}\right)+\left(\frac{1}{m_q^2}-\frac{1}{m_{\bar{q}}^2}\right)V_4\right],
\end{eqnarray}
\begin{eqnarray}
H_{anti} &=& \frac{{\boldsymbol{S}}_- \cdot {\boldsymbol{L}}}{2}\left[\left(\frac{1}{2m_q^2}-\frac{1}{2m_{\bar{q}}^2}\right) \left(\frac{1}{r}\frac{dV_c}{dr}+\frac{2}{r}\frac{dV_1}{dr}\right)\right. \nonumber \\
&& \left.+\left(\frac{1}{m_q^2}+\frac{1}{m_{\bar{q}}^2}+\frac{2}{m_qm_{\bar{q}}}\right)V_4\right].
\end{eqnarray}
The antisymmetric part $H_{anti}$ gives rise to the
the spin-orbit mixing of the heavy-light mesons with different total spins but with the same total angular momentum such as
$B(n{}^3L_L)$ and $B(n{}^1L_L)$.  Hence, the two physical states $B_L(nL)$ and $B_L^\prime(nL)$ can be
expressed as

\begin{equation}
\left(
\begin{array}{cr}
B_L(nL)\\
B^\prime_L(nL)
\end{array}
\right)
 =\left(
 \begin{array}{cr}
\cos \theta_{nL} & \sin \theta_{nL} \\
-\sin \theta_{nL} & \cos \theta_{nL}
\end{array}
\right)
\left(\begin{array}{cr}
B(n^1L_L)\\
B(n^3L_L)
\end{array}
\right),
\label{Bmixing1}
\end{equation}
where the $\theta_{nL}$ is the mixing angles.

With above formalisms, one can solve the Schr\"{o}dinger equation with Hamiltonian  of Eq.~(\ref{ha}) to get the meson wave functions, which will act as the input for calculating the strong decays of excited bottom mesons in the $^3P_0$ model.

\subsection{$^3P_0$ model}
The $^3P_0$ model was proposed by Micu \cite{Micu:1968mk} and further developed by Le Yaouanc
\cite{LeYaouanc:1972vsx, LeYaouanc:1974cvx, LeYaouanc:1977fsz,LeYaouanc:1977gm},  and it has been widely used to calculate the OZI allowed decay processes \cite{Roberts:1992js,Blundell:1996as,Barnes:1996ff,
Close:2005se,Barnes:2005pb,Zhang:2006yj,Li:2008mza,
Li:2009rka,Li:2010vx,Lu:2014zua,Pan:2016bac,Lu:2016bbk,Li:2022ybj,Li:2021qgz,Wang:2017pxm,Hao:2020fs}.
In this model, the meson decay occurs through the regroupment between the $q\bar{q}$ of the initial meson and the another $q\bar{q}$ pair created from vacuum with the quantum numbers $J^{PC}=0^{++}$. The transition operator $T$ of the decay  $A\rightarrow BC$ in the $^3P_0$ model is given by
\begin{eqnarray}
&T=-3\gamma \sum \limits_{m} \langle1m;1-m|00\rangle \int d^3\boldsymbol{p}_3 d^3\boldsymbol{p}_4 \delta^3(\boldsymbol{p}_3+\boldsymbol{p}_4)\nonumber\\
&\mathcal{Y}_{1m}\left(\frac{\boldsymbol{p}_3-\boldsymbol{p}_4}{2}\right) \chi^{34}_{1,-m} \phi^{34}_{0} \left(\omega^{34}_{0}\right)b^{\dag}_{3}(\boldsymbol{p}_3) d^{\dag}_{4}(\boldsymbol{p}_4),
\end{eqnarray}
where ${\cal{Y}}^m_1(\boldsymbol{p})\equiv|\boldsymbol{p}|^1Y^m_1(\theta_p,\phi_p)$ is solid harmonic polynomial in the momentum space of the created quark-antiquark pair.
$\chi^{34}_{1, -m}$, $\phi^{34}_0$ and $\omega^{34}_0$
 are the spin, flavor and color wave functions, respectively.
The paramtere $\gamma$ is the quark pair creation strength parameter  for $u \bar{u}$ and $d \bar{d}$ pairs, and for $s\bar{s}$ we take $\gamma_{s\bar{s}}=\gamma\frac{m_u}{m_s}$ \cite{LeYaouanc:1977gm}. The parameter $\gamma$ can be determined by fitting to the experimental data. 
The partial wave amplitude ${\cal{M}}^{LS}(\boldsymbol{P})$ of the decay  $A\rightarrow BC$ is be given by Ref.~\cite{Jacob:1959at},
\begin{eqnarray}
    {\cal{M}}^{LS}(\boldsymbol{P})&=&
    \sum_{ M_{J_B},M_{J_C}, M_S,M_L}
    \langle LM_LSM_S|J_AM_{J_A}\rangle \nonumber\\
    &&\langle
    J_BM_{J_B}J_CM_{J_C}|SM_S\rangle\nonumber\\
    &&\int
    d\Omega\,\mbox{}Y^\ast_{LM_L} 
  {\cal{M}}^{M_{J_A}M_{J_B}M_{J_C}}
    (\boldsymbol{P}),
\end{eqnarray}
where ${\cal{M}}^{M_{J_A}M_{J_B}M_{J_C}}
(\boldsymbol{P})$ is the helicity amplitude,
\begin{eqnarray}
\langle
BC|T|A\rangle =\delta^3(\boldsymbol{P}_A-\boldsymbol{P}_B-\boldsymbol{P}_C)\nonumber\\
               {\cal{M}}^{M_{J_A}M_{J_B}M_{J_C}}(\boldsymbol{P}).
\end{eqnarray}
Here, $|A\rangle$, $|B\rangle$, and $|C\rangle$ denote the mock meson states which are defined in Ref.~\cite{Hayne:1981zy}.
Then, the decay width
$\Gamma(A\rightarrow BC)$ can be expressed as

\begin{eqnarray}
\Gamma(A\rightarrow BC)= \frac{\pi
P}{4M^2_A}\sum_{LS}|{\cal{M}}^{LS}(\boldsymbol{P})|^2,
\end{eqnarray}
where $P=|\boldsymbol{P}|=\frac{\sqrt{[M^2_A-(M_B+M_C)^2][M^2_A-(M_B-M_C)^2]}}{2M_A}$,
 $M_A$, $M_B$, and $M_C$ are the masses of the mesons $A$, $B$,
and $C$, respectively. The spatial wave functions of the mesons in the $^3P_0$ model are obtained by solving the Schr$\ddot{o}$dinger equation in Eq. (\ref{ha}).

\begin{table*}[!htbp]
\begin{center}
\caption{ \label{bmass} The mass spectrum of the  bottom mesons by
different quark models in the units of MeV. The mixing angles of $B_L-B^\prime_L$ calculated in
this work are $\theta_{1P} =-53.5^\circ$, $\theta_{2P} =-54.5^\circ$, $\theta_{1D}=-50.5^\circ$, $\theta_{2D}=-50.5^\circ$ and  $\theta_{1F}=49.0^\circ$}.
\footnotesize
\begin{tabular}{lccccccccccc}
\hline\hline
  State          &PDG                 & Ours    & GI\cite{Godfrey:2016nwn}    & ARM\cite{Godfrey:2016nwn}   & NRQM\cite{Lu:2016bbk} & EFG\cite{Ebert:2009ua}  &DE\cite{DiPierro:2001dwf}  & LNR\cite{Lahde:1999ih} &ZVR\cite{Zeng:1994vj} \\\hline
  $B(1^1S_0)$  &$5279.65\pm0.12$      &5279           &5312          & 5275     & 5280   & 5280 &5279  &5277 &5280  \\
  $B(1^3S_1)$  &$5324.70\pm0.21$      &5327           &5371          & 5316     & 5329   & 5326 &5324  &5325 &5330  \\
  $B(2^1S_0)$  &$5863\pm9$            &5870           &5904          & 5834     & 5910   & 5890 &5886  &5822 &5830  \\
  $B(2^3S_1)$  &                      &5896           & 5933         & 5864     & 5939   & 5906 &5920  &5848 &5870  \\
  $B(3^1S_0)$  &                      &6278           & 6335         & 6216     & 6369   & 6379 &6320  &6117 &6210  \\
  $B(3^3S_1)$  &                      &6297           & 6355         & 6240     & 6391   & 6387 &6347  &6136 &6240  \\
  $B(1^3P_0)$  &                      &5683           & 5756         & 5720     & 5683   & 5749 &5706  &5678 &5650  \\
  $B_1(1P)$    &                      &5722           & 5777         & 5738     & 5729   & 5774 &5700  &5686 &5690  \\
  $B^\prime_1(1P)$  & $5726.1\pm1.3$  &5725           & 5784         & 5753     & 5754   & 5723 &5742  &5699 &5690  \\
  $B(1^3P_2)$  &$5739.5\pm0.7$        &5736           & 5797         & 5754     & 5768   & 5741 &5714  &5704 &5710  \\
  $B(2^3P_0)$  &                      &6104           & 6213         & 6106     & 6145   & 6221 &6163  &6010 &6060  \\
  $B_1(2P)$    &                      &6139           & 6197         & 6126     & 6185   & 6281 &6175  &6022 &6100  \\
  $B^\prime_1(2P)$   &                &6162           & 6228         & 6132     &6241    & 6209 &6194  &6028 &6100  \\
  $B(2^3P_2)$  &                      &6174           & 6213         & 6141     & 6253   & 6260 &6188  &6040 &6120  \\
  $B(1^3D_1)$  &                      &6066           & 6110         & 6053     & 6095   & 6119 &6025  &6005 &5970  \\
  $B_2(1D)$    &                      &5952           & 6095         & 6012     &6004    & 6121 &5985  &5920 &5960  \\
  $B^\prime_2(1D)$  &                 &6080           & 6124         & 6072     & 6113   & 6103 &6037  &5955 &5980  \\
  $B(1^3D_3)$   & $5971\pm5$          &5959           & 6106         & 6026     & 6014   & 6091 &5993  &5871 &5970  \\
  $B(2^3D_1)$  &                      &6420           & 6475         &6357      & 6497   & 6534 &      &6248 &      \\
  $B_2(2D)$   &                       &6334           &6450          &6334      & 6435   & 6554 &      &6179 &6310  \\
  $B^\prime_2(2D)$      &             &6433           & 6486         &6377      & 6513   & 6528 &      &6207 &6320  \\
  $B(2^3D_3)$  &                      &6341           & 6460         & 6347     & 6444   & 6542 &      &6140 &6320  \\
  $B(1^3F_2)$  &                      &6329           & 6387         &6302      & 6383   & 6412 &6264  &     &6190  \\
  $B_3(1F)$    &                      &6157           & 6358         & 6231     & 6236   & 6420 &6220  &     &6180  \\
  $B^\prime_3(1F)$       &            &6337           &6396          &6316      & 6393   & 6391 &6271  &     &6200  \\
  $B(1^3F_4)$            &            &6162           & 6364         & 6244     & 6243   & 6380 &6226  &     &6180  \\
  \hline\hline

\end{tabular}
\end{center}
\end{table*}

\section{Results and discussions}
\label{sec:result}
In the modified nonrelativistic quark model, we  determine two parameters,  $C_{q\bar{q}}=0.0152$~GeV of Eq.~(\ref{ha}) and $\beta=0.0246$~GeV of Eq.~(\ref{eq:screened}), by fitting the  masses of the bottom mesons $B$ and $B^*$, which have already been well established as the $B(1^1S_0)$ and $B(1^3S_1)$, respectively.
With these values, we calculated the mass spectrum of the bottom mesons, as shown in Table~\ref{bmass}, where we also present the predictions with other models without considering the screening effect~\cite{Zeng:1994vj,Lahde:1999ih,DiPierro:2001dwf,Ebert:2009ua,Godfrey:2016nwn, Lu:2016bbk}. It should be stressed that the mixing angles are obtained as $\theta_{1P} =-53.5^\circ$, $\theta_{2P} =-54.5^\circ$, $\theta_{1D}=-50.5^\circ$, $\theta_{2D}=-50.5^\circ$ and  $\theta_{1F}=-49.0^\circ$  by solving the potential model with the Hamiltonian  of Eq.~(\ref{ha}), which are close to the heavy quark limit mixing angles $\theta_{1P}=\theta_{2P}=-54.7^\circ$, $\theta_{1D}=\theta_{2D}=-50.8^\circ$ and $\theta_{1F}=\theta_{2F}=-49.1^\circ$ ~\cite{Cahn:2003cw}.

%

\begin{table}[!htbp]
\begin{center}
\caption{ \label{B13P2} Decay widths of the $B_2^*(5747)$ as the $B(1^3P_2)$ in units of MeV. }
\footnotesize
\begin{tabular}{lc}
\hline\hline
                               &$B_2^*(5747)$  \\\hline
  $B^+\pi^-$                   &8.2       \\
  $B^0\pi^0$                   &4.1         \\
  $B^{*+}\pi^-$                &7.8       \\
  $B^{*0}\pi^0$                &4.0             \\
  Total width                  &24.2          \\
  Experiment                   & $24.2\pm1.7$                                      \\
\hline\hline
\end{tabular}
\end{center}
\end{table}

In Table~\ref{bmass}, one can find that our results for excited bottom mesons are lower than those of Godfrey-Isgur (GI)~\cite{Godfrey:2016nwn} and Nonrelativistic quark model (NRQM)~\cite{Lu:2016bbk}, which is due to the screening effect. In the modified nonrelativistic quark model, the $B^*_2(5747)$ mass is consistent with the predicted mass of  $B(1^3P_2)$, and the $B_1(5721)$ mass is also in well agreement with the one of $B^\prime_1(1P)$.
On the other hand, the $B_J(5840)$ can be considered as the candidate of the $B(2^1S_0)$ taking into account the experimental uncertainties,  and the $B_J(5970)$ can be assigned as the candidate of the $B_2(1D)$ or $B(1^3D_3)$. As we known, only the mass information is not enough to make those assignments,  and we will calculate their strong decay widths based on these preliminary assignments for the $B_1(5721)$, $B^*_2(5747)$, $B_J(5840)$, and  $B_J(5970)$ to further examine these  assignments. 
 
 \begin{table}[!htbp]
\begin{center}
\caption{ \label{B1P} Decay widths of the $B_1(5721)$ a,s the $B(1P)$ and  $B^\prime_1(1P)$ in units of MeV, with the mixing angle $\theta_{1P}=-53.5^\circ$.}
\footnotesize
\begin{tabular}{lcc}
\hline\hline
 Channel            & $B_1(1P)$         & $B^\prime_1(1P)$        \\\hline
  $B^{*+}\pi^-$     &$129.2$            &$11.0$             \\
  $B^{*0}\pi^0$     &$64.3$             &$5.6$            \\
  Total width       &$193.5$            &$16.5$            \\
  Experiment          & \multicolumn{2}{c}{$27.5\pm3.4$} \\
\hline\hline
\end{tabular}
\end{center}
\end{table}

Before presenting the strong decays, we need to determine the quark pair creation strength $\gamma$ firstly. As we discussed above, $B_1(5721)$ may be a mixing state, and the assignments of the $B_J(5840)$ and $B_J(5970)$ are still in debate due to the unknown quantum numbers and poor decay information. In this work we take $\gamma=0.411$ by fitting to the experimental widths of $B^*_2(5747)$ which is regarded as the $B(1^3P_2)$ state in previous works~\cite{Lu:2016bbk,Kher:2017mky, Asghar:2018tha, Godfrey:2019cmi, Yu:2019iwm}. The decay properties of the state $B^*_2(5747)$ with the assignment of $B(1^3P_2)$ are shown in Table~\ref{B13P2}. The ratio of the decay modes are calculated as,
\begin{equation}
\frac{\Gamma(B^*_2(5747)\to B^{*+}\pi^-)}{\Gamma(B^*_2(5747)\to B^{+}\pi^-)}=0.95,
\end{equation}
which is consistent with the experimental data of $1.10\pm0.42\pm0.31$~\cite{D0:2007vzd} and $0.71\pm 0.14 \pm 0.30$~\cite{LHCb:2015aaf}, and also the prediction of  the nonrelativistic quark model~\cite{Lu:2016bbk}.

\begin{table}[!htbp]
\begin{center}
\caption{ \label{B5840} Decay widths of the $B_J(5840)$ as the $B(2^1S_0)$ in units of MeV.}
\scriptsize
\begin{tabular}{lc}
\hline\hline
                              &$2^1S_0$           \\\hline
  $B^{*+}\pi^-$               &$73.0$                 \\
  $B^{*0}\pi^0$               &$36.5$                   \\
  $B^*(1^3P_0)^+\pi^-$        &$0.004$                    \\
  $B^*(1^3P_0)^0\pi^0$        &$0.003$                \\
  Total width                 & $109.5$                 \\
  Experiment                &$127\pm40$ \\
\hline\hline
\end{tabular}
\end{center}
\end{table}

\begin{figure}[!htpb]
  \centering
   \begin{tabular}{c}
  \includegraphics[scale=0.7]{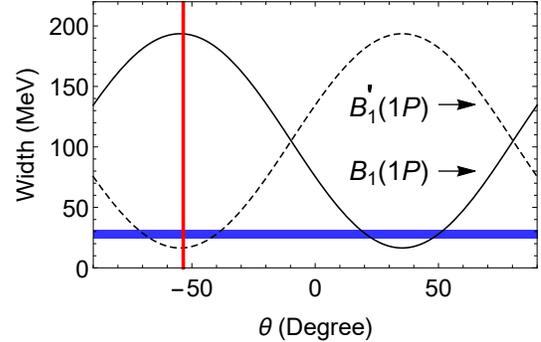}
   \end{tabular}
  \caption{Total decay width of the $B_1(5721)$ as the $B^\prime_1(1P)$ depends on the mixing angle. The vertical red solid line corresponds to the mixing angle $\theta_{1P}=-53.5^\circ$, and the blue band denotes the experimental width of the $B_1(5721)$ from RPP~\cite{PDG2021}.}
  \label{fig1}
\end{figure}

The decay widths of the $B_1(5721)$ as the $B_1(1P)$ and $B^\prime_1(1P)$ with the mixing angle $\theta_{1P}=-53.5^\circ$ are listed in Table~\ref{B1P}. The total widths of the $B_1(5721)$ as the $B_1(1P)$ and $B^\prime_1(1P)$  are predicted to be $193.5$~MeV and $16.5$~MeV, respectively.
The dependence of the total decay widths of the $B_1(5721)$ as the $B_1(1P)$ and $B^\prime_1(1P)$ 
on the mixing angle is shown in Fig.~\ref{fig1}. One can see that the total width is narrow around $\theta_{1P}=-53.5^\circ$.  We can safely rule out the $B_1(1P)$ assignment since the width of this case is much larger than the experimental data. Within in the experimental uncertainties, the total width of the $B^\prime_1(1P)$  assignment is in fair agreement with the experimental data, which implies that $B_1(5721)$ could be the $B^\prime_1(1P)$ state.

 In the heavy quark limit, the $P$-wave heavy-light mesons could be divided into $j=1/2 (0^+,1^+)$ doublet and  $j=3/2 (1^+,2^+)$ doublet, with $j(=S_q+L)$ is the total angular momentum of the light quark. For the bottom mesons, the decay width  is broad for $j=1/2$ doublet, which couples to $B\pi$ in $S$-wave, and narrow for $j=3/2$ doublet, which couples to $B\pi$ in $D$-wave. Thus, the $B_1(5721)$
and $B^*_2(5747)$ should be the  $j=3/2 (1^+,2^+)$ doublet.

%

\begin{table}[!htbp]
\begin{center}
\caption{ \label{B1D} Decay widths of the $B_J(5970)$ as the  $B_2(1D)$ and $B(1^3D_3)$ in units of  MeV with the mixing angle  $\theta_{1D}=-50.5^\circ$. }
\scriptsize
\begin{tabular}{lcc}
\hline\hline

                             &$B_2(1D)$    &$B(1^3D_3)$ \\\hline
  $B^+\pi^-$                 & $-$                     & $12.6$   \\
  $B^0\pi^0$                 & $-$                     & $6.3$   \\
  $B^{*+}\pi^-$              & $52.9$              & $13.5$   \\
  $B^{*0}\pi^0$              & $26.4$               & $6.8$     \\
  $B^*(1^3P_0)^+\pi^-$       & $0.006$               & $-$  \\
  $B^*(1^3P_0)^0\pi^0$       & $0.003$            & $-$       \\
  $B^*(1^3P_2)^+\pi^-$       & $66.8$               & $0.2$  \\
  $B^*(1^3P_2)^0\pi^0$       & $33.5$              & $0.1$   \\
  $B_1(1P)^+\pi^-$           & $0.005$            & $0.05$  \\
  $B_1(1P)^0\pi^0$           & $0.003$              & $0.03$       \\
  $B^\prime_1(1P)^+\pi^-$    & $0.4$              & $0.05$   \\
  $B^\prime_1(1P)^0\pi^0$    & $0.1$               & $0.03$       \\
  $B^0\eta$                  & $-$                  & $0.3$     \\
  $B^{*0}\eta$               & $11.2$               & $0.1$     \\
  $B_s^0K^0$                 & $-$                   & $0.1$   \\
  $B_s^{*0}K^0$              & $11.1$              & $0.03$      \\
  Total width                & $202.5$           &$40.1$       \\
  Experiment                 &\multicolumn{2}{c} {$81\pm12$ }            \\
\hline\hline
\end{tabular}
\end{center}
\end{table}

\begin{figure}[!htpb]
  \centering
   \begin{tabular}{c}
  \includegraphics[scale=0.7]{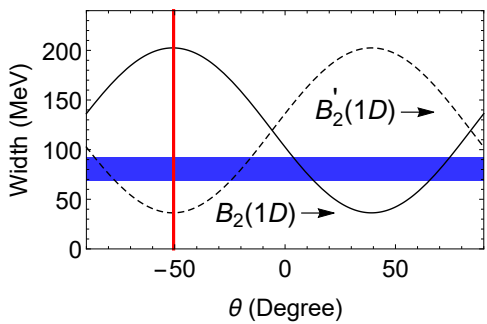}
   \end{tabular}
  \caption{Total decay width of the $B_J(5970)$    depends on the mixing angle as the $B_2(1D)$ and $B^\prime_2(1D)$. The vertical red solid line corresponds to the mixing angle $\theta_{1D}=-50.5^\circ$ and the blue band denotes the experimental width from RPP~\cite{PDG2021}}
  \label{fig2}
\end{figure}

The decay widths of the $B_J(5840)$ as the $B(2^1S_0)$ is shown in Table~\ref{B5840}, and the predicted total decay width is 109.5~MeV,  in well agreement with the experimental measurement $127\pm40$~MeV~\cite{PDG2021}. In this case, the dominant decay mode is $B^*\pi$, and the decay mode $B\pi$ is forbidden for the $B(2^1S_0)$ assignment. It should be pointed out that the decay mode $B\pi$ has not yet finally been confirmed by the LHCb~\cite{LHCb:2015aaf}, which implies that assignments of the $B_J(5840)$ as the $B(2^1S_0)$ is acceptable. 

The decay widths of the $B_J(5970)$ as the $B_2(1D)$ and $B(1^3D_3)$ are listed in Table~\ref{B1D}. The predicted total width of the $B_2(1D)$ assignment is $202.5$~MeV with mixing angle $\theta_{1D}=-50.5^\circ$, which is about 100~MeV larger than the experimental data $81\pm12$ MeV, while the one of the $B(1^3D_3)$ assignment is $40.1$ MeV. 
 The dominant decay modes are $B\pi$ and $B^*\pi$ decay modes, supported by the measurements of the CDF~\cite{CDF:2013www} and LHCb Collaborations~\cite{LHCb:2015aaf}.
The dependence of the decay widths of the $B_J(5970)$ as the $B_2(1D)$ on the mixing angle is shown in Fig.~\ref{fig2}, one can find that the total width of $B_2(1D)$ is about 200~MeV around the mixing angle $\theta_{1D}=-50.5^\circ$, and the total width of $B'_2(1D)$ is predicted to be about 50~MeV. Considering the predictive power of the model and the experimental uncertainties, it is reasonable to regard $B_J(5970)$ as the $B(1^3D_3)$.  

\section{Summary}
\label{sec:summary}
In this paper, we have calculated the bottom meson spectrum with a modified nonrelativistic quark model involving the screening effect. We present a good description of mass spectrum of the bottom mesons, especially for the excited bottom mesons. Furthermore, we also investigate the strong decay properties of the $B_1(5721)$, $B^*_2(5747)$, $B_J(5840)$, and $B_J(5970)$ with the $^3P_0$ model. 

Based on the mass spectrum and decay properties, the $B_1(5721)$ and $B^*_2(5747)$ can be identified as the $B_1^\prime(1P)$ and $B(1^3P_2)$, respectively. 
The $B_J(5840)$ could be interpreted as the $B(2^1S_0)$, and the $B_J(5970)$ could be explained as the $B(1^3D_3)$.  Further experimental information, especially the quantum numbers and decay modes of $B_J(5840)$ and $B_J(5970)$, are necessary to confirm these assignments.

\begin{acknowledgements}
 This work is supported by the Academic Improvement Project of Zhengzhou University.
\end{acknowledgements}


\begin{thebibliography}{99}

\bibitem{Chen:2016spr}
H.~X.~Chen, W.~Chen, X.~Liu, Y.~R.~Liu and S.~L.~Zhu,
A review of the open charm and open bottom systems,
Rept. Prog. Phys. \textbf{80} (2017) no.7, 076201

\bibitem{Du:2017zvv}
M.~L.~Du, M.~Albaladejo, P.~Fern\'andez-Soler, F.~K.~Guo, C.~Hanhart, U.~G.~Mei\ss{}ner, J.~Nieves and D.~L.~Yao,
Towards a new paradigm for heavy-light meson spectroscopy,
Phys. Rev. D \textbf{98} (2018) no.9, 094018


\bibitem{CLEO:1983mma}
S.~Behrends \textit{et al.} [CLEO],
Observation of Exclusive Decay Modes of $B$ Flavored Mesons,
Phys. Rev. Lett. \textbf{50} (1983), 881-884

\bibitem{PDG2021}
P.~A.~Zyla \textit{et al.} [Particle Data Group],
Review of Particle Physics,
PTEP \textbf{2020} (2020) no.8, 083C01


\bibitem{OPAL:1994hqv}
R.~Akers \textit{et al.} [OPAL],
Observations of $\pi^- B$ charge - flavor correlations and resonant $B \pi$ and $B K$ production,
Z. Phys. C \textbf{66} (1995), 19-30









\bibitem{D0:2007vzd}
V.~M.~Abazov \textit{et al.} [D0],
Observation and Properties of $L = 1 B_{1}$ and $B^*_2$ Mesons,
Phys. Rev. Lett. \textbf{99} (2007), 172001

\bibitem{CDF:2008qzb}
T.~Aaltonen \textit{et al.} [CDF],
Measurement of Resonance Parameters of Orbitally Excited Narrow $B^0$ Mesons,
Phys. Rev. Lett. \textbf{102} (2009), 102003

\bibitem{CDF:2013www}
T.~A.~Aaltonen \textit{et al.} [CDF],
Study of Orbitally Excited $B$ Mesons and Evidence for a New $B\pi$ Resonance,
Phys. Rev. D \textbf{90} (2014) no.1, 012013

\bibitem{LHCb:2015aaf}
R.~Aaij \textit{et al.} [LHCb],
Precise measurements of the properties of the $B_1(5721)^{0,+}$ and $B^\ast_2(5747)^{0,+}$ states and observation of $B^{+,0}\pi^{-,+}$ mass structures,
JHEP \textbf{04} (2015), 024










\bibitem{Zeng:1994vj}
J.~Zeng, J.~W.~Van Orden and W.~Roberts,
Heavy mesons in a relativistic model,
Phys. Rev. D \textbf{52} (1995), 5229-5241

\bibitem{Godfrey:1985xj}
S.~Godfrey and N.~Isgur,
Mesons in a Relativized Quark Model with Chromodynamics,
Phys. Rev. D \textbf{32} (1985), 189-231


\bibitem{Lahde:1999ih}
T.~A.~Lahde, C.~J.~Nyfalt and D.~O.~Riska,
Spectra and $M_1$ decay widths of heavy light mesons,
Nucl. Phys. A \textbf{674} (2000), 141-167

\bibitem{DiPierro:2001dwf}
M.~Di Pierro and E.~Eichten,
Excited Heavy - Light Systems and Hadronic Transitions,
Phys. Rev. D \textbf{64} (2001), 114004

\bibitem{Zhong:2008kd}
X.~h.~Zhong and Q.~Zhao,
Strong decays of heavy-light mesons in a chiral quark model,
Phys. Rev. D \textbf{78} (2008), 014029


\bibitem{Ebert:2009ua}
D.~Ebert, R.~N.~Faustov and V.~O.~Galkin,
Heavy-light meson spectroscopy and Regge trajectories in the relativistic quark model,
Eur. Phys. J. C \textbf{66} (2010), 197-206




\bibitem{Godfrey:2016nwn}
S.~Godfrey, K.~Moats and E.~S.~Swanson,
$B$ and $B_s$ Meson Spectroscopy,
Phys. Rev. D \textbf{94}, 054025 (2016)




 \bibitem{Lu:2016bbk}
   Q.~F.~L\"{u}, T.~T.~Pan, Y.~Y.~Wang, E.~Wang, and D.~M.~Li,
 Excited bottom and bottom-strange mesons in the quark model,
Phys.\ Rev.\ D {\bf 94}, 074012 (2016).
      
\bibitem{Kher:2017mky}
V.~Kher, N.~Devlani and A.~K.~Rai,
Spectroscopy, Decay properties and Regge trajectories of the $B$ and $B_s$ mesons,
Chin. Phys. C \textbf{41}, no.9, 093101 (2017).
      

\bibitem{Asghar:2018tha}
I.~Asghar, B.~Masud, E.~S.~Swanson, F.~Akram and M.~Atif Sultan,
Decays and spectrum of bottom and bottom strange mesons,
Eur. Phys. J. A \textbf{54} (2018) no.7, 127

\bibitem{Godfrey:2019cmi}
S.~Godfrey and K.~Moats,
Spectroscopic Assignments of the Excited $B$-Mesons,
Eur. Phys. J. A \textbf{55}, no.5, 84 (2019)

\bibitem{Yu:2019iwm}
G.~L.~Yu and Z.~G.~Wang,
Analysis of the excited bottom and bottom-strange states $B_{1}(5721)$, $B_{2}^{*}(5747)$, $B_{s1}(5830)$, $B_{s2}^{*}(5840)$, $B_{J}(5840)$ and $B_{J}(5970)$ in $B$ meson family,
Chin. Phys. C \textbf{44},  033103 (2020)



\bibitem{Chen:2022fye}
B.~Chen, S.~Q.~Luo, K.~W.~Wei and X.~Liu,
$b$-hadron spectroscopy study based on the similarity of double bottom baryon and bottom meson,
Phys. Rev. D \textbf{105} (2022) no.7, 074014

\bibitem{li:2021hss}
Q.~li, R.~H.~Ni and X.~H.~Zhong,
Towards establishing an abundant $B$ and $B_s$ spectrum up to the second orbital excitations,
Phys. Rev. D \textbf{103} (2021), 116010

\bibitem{Narison:2020wql}
S.~Narison,
Spectra and decay constants of $B_c$-like and $B_0^*$ mesons in QCD,
Phys. Lett. B \textbf{807} (2020), 135522


\bibitem{bai:2009}
B.~Q.~Li and K.~T.~Chao,
Higher Charmonia and $X$, $Y$, $Z$ states with Screened Potential,
Phys. Rev. D \textbf{79} (2009), 094004


\bibitem{Song:2015nia}
Q.~T.~Song, D.~Y.~Chen, X.~Liu and T.~Matsuki,
Charmed-strange mesons revisited: mass spectra and strong decays,
Phys. Rev. D \textbf{91} (2015), 054031

\bibitem{Song:2015fha}
Q.~T.~Song, D.~Y.~Chen, X.~Liu and T.~Matsuki,
Higher radial and orbital excitations in the charmed meson family,
Phys. Rev. D \textbf{92} (2015) no.7, 074011

\bibitem{Wang:2018rjg}
J.~Z.~Wang, Z.~F.~Sun, X.~Liu and T.~Matsuki,
Higher bottomonium zoo,
Eur. Phys. J. C \textbf{78} (2018) no.11, 915
 
\bibitem{Lakhina:2006fy}
O.~Lakhina and E.~S.~Swanson,
A Canonical $D_s(2317)$?,
Phys. Lett. B \textbf{650} (2007), 159-165


\bibitem{Li:2010vx}
D.~M.~Li, P.~F.~Ji and B.~Ma,
The newly observed open-charm states in quark model,
Eur. Phys. J. C \textbf{71} (2011), 1582


\bibitem{Wang:2019mhs}
J.~Z.~Wang, D.~Y.~Chen, X.~Liu and T.~Matsuki,
Constructing $J/\psi$ family with updated data of charmoniumlike $Y$ states,
Phys. Rev. D \textbf{99} (2019) no.11, 114003

\bibitem{Micu:1968mk}
L.~Micu,
Decay rates of meson resonances in a quark model,
Nucl. Phys. B \textbf{10} (1969), 521-526

\bibitem{LeYaouanc:1972vsx}
A.~Le Yaouanc, L.~Oliver, O.~Pene and J.~C.~Raynal,
Naive quark pair creation model of strong interaction vertices,
Phys. Rev. D \textbf{8} (1973), 2223-2234

\bibitem{LeYaouanc:1974cvx}
A.~Le Yaouanc, L.~Oliver, O.~Pene and J.~C.~Raynal,
Resonant Partial Wave Amplitudes in $\pi^+ n \rightarrow \pi^+ \pi^+ n$ According to the Naive Quark Pair Creation Model,
Phys. Rev. D \textbf{11} (1975), 1272

\bibitem{LeYaouanc:1977fsz}
A.~Le Yaouanc, L.~Oliver, O.~Pene and J.~C.~Raynal,
Strong Decays of $\psi(4.028)$ as a Radial Excitation of Charmonium,
Phys. Lett. B \textbf{71} (1977), 397-399

\bibitem{LeYaouanc:1977gm}
A.~Le Yaouanc, L.~Oliver, O.~Pene and J.~C.~Raynal,
Why Is $\psi(4.414)$ SO Narrow?,
Phys. Lett. B \textbf{72} (1977), 57-61

\bibitem{Roberts:1992js}
W.~Roberts and B.~Silvestre-Brac,
General method of calculation of any hadronic decay in the $^3P_0$ model,
Few Body Syst. \textbf{11} (1992) no.4, 171-193

\bibitem{Blundell:1996as}
  H.~G.~Blundell,
  Meson properties in the quark model: A look at some outstanding problems,
  hep-ph/9608473.


 \bibitem{Barnes:1996ff}
     T.~Barnes, F.~E.~Close, P.~R.~Page, and E.~S.~Swanson,
        Higher quarkonia,
 Phys.\ Rev.\ D {\bf 55}, 4157 (1997).


\bibitem{Close:2005se}
    F.~E.~Close and E.~S.~Swanson,
Dynamics and decay of heavy-light hadrons,
Phys.\ Rev.\ D {\bf 72}, 094004 (2005).


\bibitem{Barnes:2005pb}
    T.~Barnes, S.~Godfrey, and E.~S.~Swanson,
       Higher charmonia,
        Phys.\ Rev.\ D {\bf 72}, 054026 (2005).



\bibitem{Zhang:2006yj}
    B.~Zhang, X.~Liu, W.~Z.~Deng, and S.~L.~Zhu,
       $D_{sJ}(2860)$ and $D_{sJ}(2715)$,
        Eur.\ Phys.\ J.\ C {\bf 50}, 617 (2007).



\bibitem{Li:2008mza}
  D.~M.~Li and B.~Ma,
  $X(1835)$ and $\eta(1760)$ observed by the BES Collaboration,
  Phys.\ Rev.\ D {\bf 77}, 074004 (2008).







\bibitem{Li:2009rka}
  D.~M.~Li and E.~Wang,
  Canonical interpretation of the $\eta_2(1870)$,
  Eur.\ Phys.\ J.\ C {\bf 63},297 (2009).





\bibitem{Lu:2014zua}
    Q.~F.~L\"{u} and D.~M.~Li,
Understanding the charmed states recently observed by the LHCb and BaBar Collaborations in the quark model,
Phys.\ Rev.\ D {\bf 90}, 054024 (2014).


\bibitem{Pan:2016bac}
  T.~T.~Pan, Q.~F.~L\"{u} , E.~Wang and D.~M.~Li,
  Strong decays of the $X(2500)$ newly observed by the BESIII Collaboration,
  Phys.\ Rev.\ D {\bf 94}, no. 5, 054030 (2016).

\bibitem{Li:2022ybj}
T.~G.~Li, Z.~Gao, G.~Y.~Wang, D.~M.~Li, E.~Wang and J.~Zhu,
The possible assignments of the scalar $K_0^*(1950)$ and $K_0^*(2130)$ within the $^3P_0$ model,
[arXiv:2203.17082 [hep-ph]].

\bibitem{Li:2021qgz}
Z.~Y.~Li, D.~M.~Li, E.~Wang, W.~C.~Yan and Q.~T.~Song,
Assignments of the $Y(2040)$, $\rho(1900)$, and $\rho(2150)$ in the quark model,
Phys. Rev. D \textbf{104} (2021) no.3, 034013



\bibitem{Wang:2017pxm}
G.~Y.~Wang, S.~C.~Xue, G.~N.~Li, E.~Wang and D.~M.~Li,
Strong decays of the higher isovector scalar mesons,
Phys. Rev. D \textbf{97} (2018) no.3, 034030

\bibitem{Hao:2020fs}
W.~Hao, G.~Y.~Wang, E.~Wang, G.~N.~Li and D.~M.~Li,
Canonical interpretation of the $X(4140)$ state within the $^3P_0$ model,
Eur. Phys. J. C \textbf{80} (2020) no.7, 626

\bibitem{Jacob:1959at}
M.~Jacob and G.~C.~Wick,
On the General Theory of Collisions for Particles with Spin,
Annals Phys. \textbf{7} (1959), 404-428

\bibitem{Hayne:1981zy}
 C.~Hayne and N.~Isgur,
Beyond the Wave Function at the Origin: Some Momentum Dependent Effects in the Nonrelativistic Quark Model,
Phys. Rev. D \textbf{25} (1982), 1944


\bibitem{Cahn:2003cw}
R.~N.~Cahn and J.~D.~Jackson,
Spin orbit and tensor forces in heavy quark light quark mesons: Implications of the new $D_{(s)}$ state at 2.32-GeV,
Phys. Rev. D \textbf{68}, 037502 (2003)








\end{thebibliography}
\end{document}